\documentclass[twocolumn]{jpsj3}
\usepackage{txfonts}
\usepackage{graphicx,bm,color,braket}

\title{Ground-State Wave Function with Interactions between Different Species in $M$-Component Miscible Bose--Einstein Condensates}
\author{Wataru Kohno, Akimitsu Kirikoshi, and Takafumi Kita}
\inst{Department of Physics, Hokkaido University, Sapporo 060-0810, Japan}
\abst{
We construct a variational ground-state wave function of weakly interacting $M$-component Bose--Einstein condensates beyond the mean-field theory by incorporating the dynamical 3/2-body processes, where one of the two colliding particles drops into the condensate and vice versa.
Our numerical results with various masses and particle numbers show that the 3/2-body processes between different particles make finite contributions to lowering the ground-state energy, implying that many-body correlation effects between different particles are essential even in the weak-coupling regime of the Bose--Einstein condensates.
We also consider the stability condition for $2$-component miscible states using the new ground-state wave function.
 Through this calculation, we obtain the relation $U^{2}_{AB}/U_{AA}U_{BB}<1+\alpha$, where $U_{ij}$ is the effective contact potential between particles $i$ and $j$ and $\alpha$ is the correction, which originates from the $3/2$-body and $2$-body processes.
}

\begin{document}
\maketitle
\section{Introduction}
Multicomponent Bose--Einstein condensates (BECs) has been studied extensively since its experimental demonstrations {with trapped dilute gases} \cite{ex1,ex2,ex3,ex4,ex5}. 
In particular, many theoretical studies focusing on the collective features of condensates have been carried out, such as dynamical instabilities and collapsing processes \cite{Th1,Th2,Th3,Th4,Th5,Th6,Th6pl}, configurations of trapped condensates \cite{Th1,Th2,Th3,Th7}, and quantized vortices due to the topological defects \cite{Th8,Th9}.
On the other hand, the collisional processes smaller than $O(N_{i})$ tend to be neglected when considering the behavior of condensates of dilute gasses, where $N_{i}\gg1$ is the number of particle $i$.
Recently, the processes with the order of $O(\sqrt{N_{i}})$, such as the $2$-body process given by $NC_{i}+NC_{i}\leftrightarrow NC_{i}+NC_{i}$ \cite{GA} and the $3/2$-body process given by by $NC_{i}+NC_{i}\leftrightarrow C_{i}+NC_{i}$, where $C_{i}$ ($NC_{i}$) denotes condensate (non-condensate) $i$, have been incorporated in the variational wave function self-consistently beyond the mean-field approximation for homogeneous single-component BECs system at $T=0$ \cite{kita3/2}.
According to the results of Ref. \citen{kita3/2}, the 3/2-body processes cause finite lifetimes of quasiparticles at long wavelengths and play an essential role in maintaining the macroscopic coherent state of the BEC in equilibrium.
However, the ground-state wave function with $3/2$-body processes between different particles has never been constructed beyond the mean-field theory in Bose--Bose mixtures.

In this paper, we generalize the variational method for single-component BECs \cite{kita3/2} to a mixed system and construct the ground-state wave function by superposing the variational parameters that characterize $2$-body and $3/2$-body processes between different particles such as $NC_{i}+NC_{j}\leftrightarrow NC_{i}+NC_{j}$ and $NC_{i}+NC_{j}\leftrightarrow C_{i}+NC_{j}$, respectively.
From the self-consistent equations determined by the energy-minimum conditions, we numerically obtain the variational parameters for the $2$-component mixture (particles $A$ and $B$) with various masses ($m_{A}$ versus $m_{B}$) and particle numbers ($N_{A}$ versus $N_{B}$), where $m_{i}$ denotes the mass of particle $i$.
As described below, our numerical results show that the 2-body and 3/2-body processes between different species play roles in lowering the ground-state energy as well as those between the same species.
This implies that many-body correlations between different particles are also essential and should be incorporated in miscible multi-component systems.
{{
In addition to the evaluation of ground-state energies, we derive the correction to the stability condition for $2$-component miscible BECs including $2$-body and $3/2$-body processes. 
While the conventional condition without $2$-body and $3/2$-body processes is given by $U^{2}_{AB}/U_{AA}U_{BB}<1$, where $U_{ij}$ is the contact potential between particle $i$ and $j$, many-body effects give finite contributions to this condition as $1\to 1+\alpha$.
In this context, many-body effects in multi-component BECs may change the critical points from miscible states to other states, such as droplet-formed states\cite{Th6} $(U_{AB}<0)$ and phase-separated states\cite{Th6pl} $(U_{AB}>0)$.
}}

This paper is organized as follows.
In Sect.\ 2, we construct a variational wave function for the ground state with the $2$-body and $3/2$-body processes between different particles, obtain an expression for the ground-state energy, derive the equations to determine the energy minimum, and give the stability condition for $2$-component miscible BECs.
  Section 3 outlines numerical procedures for this analysis and presents results. 
  In particular, we present (i) the ground-state energies, (ii) the behavior of variational parameters, and (iii) the correction value $\alpha$ and its related parameters.
  Section 4 summarizes the paper.

\section{Construction of the Ground State in $M$-Component Mixed BEC}
We here describe a dilute BEC composed of $M$ types of spineless bosons with total particle number $N_{\rm{all}}\equiv N_1+N_2{+}\cdots {+}N_M$, where $N_i$ denotes the number of particle $i$.
Each mass of particles in the mixture is labeled by $m_{1},m_{2},\cdots m_{M}$.
In this section, we construct the number-conserving variational wave function for the ground state.
\subsection{Hamiltonian and number-conserving operators}
In this paper, we consider a system described by the following second-quantized Hamiltonian:

\begin{align}
\hat{H}&= \sum_{i=1}^{M}{\sum_{\bm{k}}}^{'}\varepsilon^{i}_{k}\hat{c}^{\dagger}_{i}(\bm{k})\hat{c}_{i}(\bm{k})\notag\\
&+\sum_{i,j=1}^{M}\frac{U_{ij}}{2\mathcal{V}}\sum_{\bm{k},\bm{k}',\bm{q}}\hat{c}^{\dagger}_{i}(\bm{k}+\bm{q})\hat{c}^{\dagger}_{j}(\bm{k}'-\bm{q})\hat{c}_{j}(\bm{k}')\hat{c}_{i}(\bm{k}),\label{Ham}
\end{align}
where the primed sum is defined by $\displaystyle{\sum_{\bm k}}'\equiv \sum_{\bm k}(1-\delta_{\bm{k},\bm{0}})$,  $\varepsilon^{i}_{k}\equiv \hbar^{2}k^{2}/2m_{i}$ {denotes} the kinetic energy, 
$\mathcal{V}$ is the volume of the system, and {{$U_{ij}=U_{ji}$ is the effective contact potential for treating the scattering effect between particles $i$ and $j$\cite{Pethic}}}.
Our aim is to construct the ground-state wave function of Eq.\ (\ref{Ham}) that describes the miscible state of a weakly interacting $M$-component BEC.
To carry out this, we classify $\hat{H}$ according to the number of non-condensed states involved as
\begin{equation}
\hat{H}=\hat{H}_{0}+\hat{H}_{1}+\hat{H}_{\frac{3}{2}}+\hat{H}_{2}
\end{equation}
with
 \begin{subequations}
\begin{align}
\hat{H}_{0}&\equiv
\sum_{i,j=1}^{M}\frac{U_{ij}}{2\mathcal{V}}\hat{c}^{\dagger}_{i}(\bm{0})\hat{c}^{\dagger}_{j}(\bm{0})\hat{c}_{j}(\bm{0})\hat{c}_{i}(\bm{0}){,} \\
\hat{H}_{1}&\equiv \sum_{i=1}^{M}{\sum_{\bm{k}}}^{'}\varepsilon^{i}_{k}\hat{c}^{\dagger}_{i}(\bm{k})\hat{c}_{i}(\bm{k})\notag\\
&+\sum_{i,j=1}^{M}\frac{U_{ij}}{\mathcal{V}}{\sum_{\bm{k}}}^{'}\left[\hat{c}^{\dagger}_{i}(\bm{0})\hat{c}^{\dagger}_{j}(\bm{k})\left\{\hat{c}_{j}(\bm{k})\hat{c}_{i}(\bm{0})+\hat{c}_{j}(\bm{0})\hat{c}_{i}(\bm{k})\right\}\right]\notag\\
&+\sum_{i,j=1}^{M}\frac{U_{ij}}{2\mathcal{V}}{\sum_{\bm{k}}}^{'}\left[\hat{c}^{\dagger}_{i}(\bm{0})\hat{c}^{\dagger}_{j}(\bm{0})\hat{c}_{j}(\bm{k})\hat{c}_{i}(-\bm{k})+{\rm{H.C}}\right] {,}\\
\hat{H}_{\frac{3}{2}}&\equiv \sum_{i,j=1}^{M}\frac{U_{ij}}{\mathcal{V}}{\sum_{\bm{k}_{1},\bm{k}_{2},\bm{k}_{3}}}^{'}\delta_{\bm{k}_{1}+\bm{k}_{2}+\bm{k}_{3},\bm{0}}\notag\\
&\times\left[\hat{c}^{\dagger}_{i}(\bm{0})\hat{c}^{\dagger}_{j}(-\bm{k}_{3})\hat{c}_{j}(\bm{k}_{2})\hat{c}_{i}(\bm{k}_{1})+{\rm{H.C}}\right]\label{3/2body} {,}\\
\hat{H}_{2}&\equiv \sum_{i,j=1}^{M}\frac{U_{ij}}{2\mathcal{V}}{\sum_{\bm{k},\bm{k}',\bm{q}}}^{'}\hat{c}^{\dagger}_{i}(\bm{k}+\bm{q})\hat{c}^{\dagger}_{j}(\bm{k}'-\bm{q})\hat{c}_{j}(\bm{k}')\hat{c}_{i}(\bm{k}),\label{2body}
\end{align}\label{Hamiltonian}
\end{subequations}
where H.C{.}\ denotes the Hermitian conjugate.

Next, we introduce the number-conserving creation-annihilation operators \cite{kita3/2,GA2} to consider the number-conserved systems.
First, orthonormal basis functions for $\bm{k}=\bm{0}$ (the condensate) are given by 
\begin{align}
&\ket{n_{1},n_{2},\cdots,n_{M}}_{0}\notag\\
&{{\equiv \frac{(\hat{c}_{1}^{\dagger}(\bm{0}))^{n_{1}}}{\sqrt{n_{1}!}}\ket{0}_{1}\otimes\frac{(\hat{c}_{2}^{\dagger}(\bm{0}))^{n_{2}}}{\sqrt{n_{2}!}}\ket{0}_{2}\cdots\otimes\frac{(\hat{c}_{M}^{\dagger}(\bm{0}))^{n_{M}}}{\sqrt{n_{M}!}}\ket{0}_{M},}}
\end{align}
where $n_{i}=0,1,2,\cdots,N_{i}$ and $\ket{0}_{i}$ denotes the vacuum state of particles $i$ with (i) $_{i}{\braket{0|0}}_{i}=1$ and (ii) $\hat{c}_{i}(\bm{k})\ket{0}=0$ for any $\bm{k}$.
The ground state without interactions is given by $\ket{N_{1},N_{2},\cdots,N_{M}}_{0}$.

Second, we introduce operators $(\hat{\beta}_{i}^{\dagger},\hat{\beta}_{i})$ for $n_{i}\geq0$, which satisfy the following relations:

\begin{subequations}
\begin{align}
&\hat{\beta}^{\dagger}_{i}\ket{n_{1},\cdots,n_{i},\cdots,n_{M}}_{0}=\ket{n_{1},\cdots,n_{i}+1,\cdots,n_{M}}_{0}, \\
&\hat{\beta}_{i}\ket{n_{1},\cdots,n_{i}+1,\cdots,n_{M}}_{0}=\ket{n_{1},\cdots,n_{i},\cdots,n_{M}}_{0},
\end{align}
\end{subequations}
with $\hat{\beta}_{i}\ket{n_{1},\cdots,n_{i}=0,\cdots,n_{M}}_{0} =0$.
These operators are expressible in terms of $\hat{c}^{\dagger}_{i}(\bm{0})$ and $\hat{c}_{i}(\bm{0})$ as {{ 
\begin{subequations}
\begin{align}
&\hat{\beta}_{i}^{\dagger}=\hat{c}^{\dagger}_{i}(\bm{0})[1+\hat{c}^{\dagger}_{i}(\bm{0})\hat{c}_{i}(\bm{0})]^{-\frac{1}{2}},\\
&\hat{\beta}_{i}=[1+\hat{c}^{\dagger}_{i}(\bm{0})\hat{c}_{i}(\bm{0})]^{-\frac{1}{2}}\hat{c}_{i}(\bm{0}),
\end{align}
\end{subequations}}}
and satisfy
\begin{subequations}
\begin{align}
&\hat{\beta}_{i}^{\nu}(\hat{\beta}_{i}^{\dagger})^{\nu}\ket{n_{1},\cdots,n_{i},\cdots,n_{M}}_{0}=\ket{n_{1},\cdots,n_{i},\cdots,n_{M}}_{0},\\
&(\hat{\beta}_{i}^{\dagger})^{\nu}\hat{\beta}_{i}^{\nu}\ket{n_{1},\cdots,n_{i},\cdots,n_{M}}_{0}\notag\\
&=\begin{cases}
 \ket{n_{1},\cdots,n_{i},\cdots,n_{M}}_{0} & (\nu\leq n_{i}) \\
  0 & (\nu> n_{i}).
 \end{cases}
\end{align}
\end{subequations}
 Hence, $(\hat{\beta}_{i}^{\dagger})^{\nu}\hat{\beta}_{i}^{\nu}\simeq1$ and $\hat{\beta}_{i}^{\nu}(\hat{\beta}_{i}^{\dagger})^{\nu}=1$  for $\nu=1,2,\cdots$.
 The former approximation holds exactly in the weak-coupling regime, where the ground state is composed of the kets $\ket{n_{1},\cdots,n_{i},\cdots,n_{M}}_{0}$ with $n_{i}=O(N_{i})= O(N_{\rm{all}})$. Hereafter we replace $\simeq$ by $=$.

Using $\hat{\beta}_{i}^{\dagger}$ and $\hat{\beta}_{i}$, we define the number-conserving creation-annihilation operators for non-condensed particles $i$ ($\bm{k}\neq \bm{0}$) by 
\begin{equation}
\hat{d}^{\dagger}_{i}(\bm{k})\equiv\hat{c}^{\dagger}_{i}(\bm{k})\hat{\beta}_{i} , \ \hat{d}_{i}(\bm{k})\equiv\hat{\beta}_{i}^{\dagger}\hat{c}_{i}(\bm{k}),
\end{equation}
where $\hat{d}^{\dagger}_{i}(\bm{k})$ and $\hat{d}_{i}(\bm{k})$ satisfy
 \begin{subequations}
 \begin{align}
[\hat{d}_{i}(\bm{k}),\hat{d}^{\dagger}_{j}(\bm{k}')]&= \delta_{i,j}\delta_{\bm{k},\bm{k}'} \label{statistic}, \\
 [\hat{d}_{i}(\bm{k}),\hat{d}_{j}(\bm{k}')]&=[\hat{d}^{\dagger}_{i}(\bm{k}),\hat{d}^{\dagger}_{j}(\bm{k}')]=0.
 \end{align}
 \end{subequations}
The introduced operator $\hat{d}^{\dagger}_{i}(\bm{k})$ has the physical meaning of exciting a particle from condensate $i$ to the excited state with $\bm{k}{\neq {\bm 0}}$.

\subsection{{{Girardeau--Arnowitt wave function for mixed condensates}}}

{{Next, we introduce the ground state that characterizes the pair-interaction processes involved in $\hat{H}_{1}$ and $\hat{H}_{2}$ such as $C+NC\leftrightarrow C+NC$, $C+C\leftrightarrow NC+NC$, and $NC+NC\leftrightarrow NC+NC$, where $C$ and $NC$ respectively denote condensates and non-condensates.
The wave function for a single-component system was given by Girardeau and Arnowitt\cite{GA}.
Here, we apply the method introduced by Ref. \citen{kita3/2}} to multi-component systems using $\hat{d}_{i}(\bm{k})$, $\hat{d}^{\dagger}_{i}(\bm{k})$, and $\ket{N_{1},N_{2},\cdots,N_{M}}_{0}$.}

First, we define the pair-creating operator $\hat{\pi}_{ij}$ by
\begin{equation}
\hat{\pi}_{ij}^{\dagger}\equiv\frac{1}{2} {\sum_{\bm{k}}}^{'}\phi_{ij}(\bm{k})\hat{c}_{i}^{\dagger}(\bm{k})\hat{c}_{j}^{\dagger}(-\bm{k}),
\end{equation}
where $\phi_{ij}(\bm{k})$ is a variational parameter what characterizes the pair excitation of particles $(i,\bm{k})$ and $(j,-\bm{k})$ from condensates caused by interactions between particles.
 Its number-conserving correspondent $\hat{\Pi}_{\rm{GA}}^{\dagger}$ is given by 
\begin{equation}
\hat{\Pi}_{\rm{GA}}^{\dagger}\equiv\sum_{i,j=1}^{M}\hat{\pi}_{ij}^{\dagger}\hat{\beta}_{i}\hat{\beta}_{j}= \frac{1}{2}\sum_{i,j=1}^{M} {\sum_{\bm{k}}}^{'}\phi_{ij}(\bm{k})\hat{d}_{i}^{\dagger}(\bm{k})\hat{d}_{j}^{\dagger}(-\bm{k}),
\end{equation}
satisfying 
\begin{equation}
[\hat{d}_{i}(\bm{k}),\hat{\Pi}_{\rm{GA}}^{\dagger}]=\sum_{j=1}^{M}\phi_{ij}(\bm{k})\hat{d}_{j}^{\dagger}(-\bm{k}).
\end{equation}
Using the operator, we can express the ground state with pair processes as 
\begin{equation}
\ket{\Phi_{\rm{GA}}}=A_{\rm{GA}}{\rm{exp}} \left(\hat{\Pi}_{\rm{GA}}^{\dagger}\right)\ket{N_{1},N_{2},\cdots,N_{M}}_{0},
\end{equation}
where $A_{\rm{GA}}$ is the normalization constant determined by $\braket{\Phi_{\rm{GA}}|\Phi_{\rm{GA}}}=1$.
Operating $\hat{d}_{i}(\bm{k})$ {on} $\ket{\Phi_{\rm{GA}}}$, we obtain
\begin{equation}
\hat{d}_{i}(\bm{k})\ket{\Phi_{\rm{GA}}}=\sum_{j=1}^{M}\phi_{ij}(\bm{k})\hat{d}_{j}^{\dagger}(-\bm{k})\ket{\Phi_{\rm{GA}}},
\end{equation}
where we used the relation $\phi_{ij}(\bm{k})=\phi_{ji}(-\bm{k})$ originating from the commutator relations of bosons.
Next, we introduce the following quantities:

\begin{align}
&\Phi_{ij}(\bm{k})\equiv \delta_{ij}-\sum_{k=1}^{M}\phi_{ik}(\bm{k})\phi^{*}_{kj}(-\bm{k})=\Phi^{*}_{ji}(\bm{k})\notag\\
&\rightarrow \underline{\Phi}(\bm{k})=\underline{1}-\underline{\phi}(\bm{k}) \underline{\phi}^{\dagger}(\bm{k}), \label{Phi}
\end{align}
\begin{subequations}
\begin{align}
&u_{ij}(\bm{k})\equiv \left(\underline{\Phi}^{-\frac{1}{2}}(\bm{k})\right)_{ij}=u^{*}_{ji}(\bm{k}),\label{u}\\
&v_{ij}(\bm{k})\equiv (\underline{u}(\bm{k})\underline{\phi}(\bm{k}))_{ij}=v_{ji}(-\bm{k}),\label{v}
\end{align}
\end{subequations}
where {$\underline{1}$ and $\underline{\phi}(\bm{k})$ are} $M\times M$ matrices defined by $(\underline{1})_{ij}=\delta_{ij}$ and 
 $(\underline{\phi}(\bm{k}))_{ij}=\phi_{ij}({\bm k})$, respectively.
 Using them, we define the following Bogoliubov operators:

\begin{subequations}
\begin{align}
&\hat{{\gamma}}_{i}(\bm{k})\equiv \sum_{j=1}^{M}[u_{ij}(\bm{k})\hat{d}_{j}(\bm{k})-v_{ij}(\bm{k})\hat{d}_{j}^{\dagger}(-\bm{k})] \label{op1}, \\
&\hat{{\gamma}}^{\dagger}_{i}(-\bm{k})\equiv\sum_{j=1}^{M}[-v^{*}_{ij}(-\bm{k})\hat{d}_{j}(\bm{k})+u^{*}_{ij}(-\bm{k})\hat{d}^{\dagger}_{j}(-\bm{k})] \label{op2}.
\end{align}
\end{subequations}
We can easily confirm the relation $[\hat{{\gamma}}_{i}(\bm{k}),\hat{{\gamma}}^{\dagger}_{i'}(\bm{k}')]=\delta_{ii'}\delta_{\bm{k}\bm{k}'}$.
The expressions for $\hat{d}$ and $\hat{d}^{\dagger}$ in terms of $\hat{{\gamma}}$ and $\hat{{\gamma}}^{\dagger}$ are given by
\begin{subequations}
\begin{align}
&\hat{d}_{i}(\bm{k})= \sum_{j=1}^{M}[u_{ij}(\bm{k})\hat{{\gamma}}_{j}(\bm{k})+v_{ij}(\bm{k})\hat{{\gamma}}_{j}^{\dagger}(-\bm{k})],\\
&-\hat{d}^{\dagger}_{i}(-\bm{k})=\sum_{j=1}^{M}[-v^{*}_{ij}(-\bm{k})\hat{{\gamma}}_{j}(\bm{k})-u^{*}_{ij}(-\bm{k})\hat{{\gamma}}^{\dagger}_{j}(-\bm{k})].
\end{align}
\end{subequations}

We note that the introduced vector $\ket{\Phi_{\rm{GA}}}$ is normalized and characterized as the vacuum of $\hat{{\gamma}}_{i}(\bm{k})$.
We can easily confirm the connection between $\ket{\Phi_{\rm{GA}}}$ and the Girardeau--Arnowitt wave function by putting $M=1$.
In other words, $\phi_{ij}(\bm{k})$ with $i\neq j$ involved in $\ket{\Phi_{\rm{GA}}}$ is the new factor, which characterizes the effects of interactions between different particles in $M$-component BECs. 
However, $\ket{\Phi_{\rm{GA}}}$ only includes pair processes via $\phi$ meaning that it has no contributions from $3/2$-body processes, i.e., $\bra{\Phi_{\rm{GA}}}\hat{H}_{\frac{3}{2}}\ket{\Phi_{\rm{GA}}}=0$.
To incorporate $3/2$-body processes, we need to characterize them by introducing the corresponding variational parameters as outlined in Sect. 2.3.

\subsection{Including 3/2 processes in mixed condensates}

We straightforwardly generalize the method in Ref. \citen{kita3/2} to mixed BECs.
To begin with, we introduce the following operator describing the 3/2 processes:

\begin{align}
&\hat{\Pi}_{\frac{3}{2}}^{\dagger}\equiv\frac{1}{3!}\sum_{i,j,k=1}^{M}{\sum_{\bm{k}_{1},\bm{k}_{2},\bm{k}_{3}}}^{'}w_{ijk}(\bm{k}_{1},\bm{k}_{2},\bm{k}_{3})\notag\\
&\times\hat{{\gamma}}^{\dagger}_{i}(\bm{k}_{1})\hat{{\gamma}}^{\dagger}_{j}(\bm{k}_{2})\hat{{\gamma}}^{\dagger}_{k}(\bm{k}_{3}),
\end{align}
where $w_{ijk}(\bm{k}_{1},\bm{k}_{2},\bm{k}_{3})$ is a variational parameter characterizing $3/2$-body processes what satisfies
$\hat{P}w_{ijk}(\bm{k}_{1},\bm{k}_{2},\bm{k}_{3})=w_{ijk}(\bm{k}_{1},\bm{k}_{2},\bm{k}_{3})$ for any permutation $\hat{P}$ with three elements {$(i,\bm{k}_{1})$, $(j,\bm{k}_{2})$, and $(k,\bm{k}_{3})$}.

Using this operator, we introduce an improved variational ket as follows:

\begin{equation}
\ket{\Phi}=A_{\frac{3}{2}}{\rm{exp}} \left(\hat{\Pi}_{\frac{3}{2}}^{\dagger}\right)\ket{\Phi_{\rm{GA}}},
\end{equation}
where $A_{\frac{3}{2}}$ is the normalization constant determined by $\braket{\Phi|\Phi}=1$.
The functional derivative of ${\rm{ln}}A^{-2}_{\frac{3}{2}}$ with respect to $w_{ijk}(\bm{k}_{1},\bm{k}_{2},\bm{k}_{3})$ yields
\begin{align}
 &\frac{\delta {\rm{ln}}A^{-2}_{\frac{3}{2}}}{\delta w_{ijk}(\bm{k}_{1},\bm{k}_{2},\bm{k}_{3})}
= A^{2}_{\frac{3}{2}} \frac{\delta A^{-2}_{\frac{3}{2}}}{\delta w_{ijk}(\bm{k}_{1},\bm{k}_{2},\bm{k}_{3})}\notag\\
& =A^{2}_{\frac{3}{2}}\bra{\Phi_{\rm{GA}}}  {\rm{exp}} \left(\hat{\Pi}_{\frac{3}{2}}\right)\frac{\delta {\rm{exp}}  \left(\hat{\Pi}_{\frac{3}{2}}^{\dagger}\right)  }{\delta w_{ijk}(\bm{k}_{1},\bm{k}_{2},\bm{k}_{3})}\ket{\Phi_{\rm{GA}}}\notag\\
&=A^{2}_{\frac{3}{2}}\bra{\Phi_{\rm{GA}}}  {\rm{exp}} \left(\hat{\Pi}_{\frac{3}{2}}\right)\frac{\delta  \hat{\Pi}_{\frac{3}{2}}^{\dagger}  }{\delta w_{ijk}(\bm{k}_{1},\bm{k}_{2},\bm{k}_{3})}{\rm{exp}}  \left(\hat{\Pi}_{\frac{3}{2}}^{\dagger}\right) \ket{\Phi_{\rm{GA}}}\notag\\
&=\bra{\Phi}\hat{{\gamma}}^{\dagger}_{i}(\bm{k}_{1})\hat{{\gamma}}^{\dagger}_{j}(\bm{k}_{2})\hat{{\gamma}}^{\dagger}_{k}(\bm{k}_{3}) \ket{\Phi}.
\label{dlnA/dw}
\end{align}
Operators $\hat{\gamma}_{i}(\bm{k})$ and $ \hat{\Pi}_{\frac{3}{2}}^{\dagger}$ also satisfy
\begin{align}
[\hat{\gamma}_{i}(\bm{k}),\hat{\Pi}_{\frac{3}{2}}^{\dagger}]
&=\frac{1}{2}{\sum_{\bm{k}_{2},\bm{k}_{3}}}' \sum_{a,b=1}^{M}w_{iab} (\bm{k},\bm{k}_{2},\bm{k}_{3})
\hat{\gamma}^{\dagger}_{a}(\bm{k}_{2})\hat{\gamma}^{\dagger}_{b}(\bm{k}_{3}) , \label{comu}
\end{align}
so $[\hat{\gamma}_{i}(\bm{k}),\exp\left(\hat{\Pi}_{\frac{3}{2}}^{\dagger}\right)]=[\hat{\gamma}_{i}(\bm{k}),\hat{\Pi}_{\frac{3}{2}}^{\dagger}]\exp\left(\hat{\Pi}_{\frac{3}{2}}^{\dagger}\right)$.

The normalization constant $A^{-2}_{\frac{3}{2}}$ is evaluated analytically in Appendix B of Ref. \citen{kita3/2} for $M=1$ systems.
Here, we apply Eq.\ (B.1) of Ref. \citen{kita3/2} to $M$-component systems as
 \begin{equation}
{\rm{ln}}A^{-2}_{\frac{3}{2}}=\sum^{\infty}_{\nu=1}\frac{J_{3\nu}}{(\nu!)^{2}}, \ J_{3\nu}\equiv\bra{\Phi}\hat{\Pi}_{\frac{3}{2}}^{\nu}(\hat{\Pi}_{\frac{3}{2}}^{\dagger})^{\nu}\ket{\Phi}_{\rm{c}},\label{lnA}
\end{equation}
where c denotes the connected subgroups of $\bra{\Phi}\hat{\Pi}_{\frac{3}{2}}^{\nu}(\hat{\Pi}_{\frac{3}{2}}^{\dagger})^{\nu}\ket{\Phi}$ when writing diagrams as shown in Fig.\ B.\ 1 of  Ref.\ \citen{kita3/2}.
{Especially} in the weak-coupling region, ${\rm{ln}}A^{-2}_{\frac{3}{2}}$ is given to a good approximation by
\begin{equation}
{\rm{ln}}A^{-2}_{\frac{3}{2}}\simeq {J_{3}=} \frac{1}{3!}\sum_{i,j,k=1}^{M}{\sum_{\bm{k}_{1},\bm{k}_{2},\bm{k}_{3}}}^{'}|w_{ijk}(\bm{k}_{1},\bm{k}_{2},\bm{k}_{3})|^{2}, \label{lnAap}
\end{equation}
as the lowest-order contribution because the terms of $\nu \geq 2$ in Eq. (\ref{lnA}), which have increasing numbers of summations over $\bm{k}\neq \bm{0}$, are negligible compared with Eq. (\ref{lnAap}).
Therefore, we obtain
\begin{equation}
{{w^{*}_{ijk}(\bm{k}_{1},\bm{k}_{2},\bm{k}_{3})}}\simeq \bra{\Phi}\hat{{\gamma}}^{\dagger}_{i}(\bm{k}_{1})\hat{{\gamma}}^{\dagger}_{j}(\bm{k}_{2})\hat{{\gamma}}^{\dagger}_{k}(\bm{k}_{3}) \ket{\Phi}\label{w}
\end{equation}
by substituting Eq.\ (\ref{lnAap}) into Eq.\ (\ref{dlnA/dw}).

For later convenience, we introduce $ \rho^{\frac{3}{2}}_{ij}(\bm{k})\equiv\bra{\Phi}\hat{{\gamma}}^{\dagger}_{i}(\bm{k})\hat{{\gamma}}_{j}(\bm{k}) \ket{\Phi}$ {{as follows:
}}
{{\begin{align}
 \rho^{\frac{3}{2}}_{ij}(\bm{k})&=\bra{\Phi}\hat{{\gamma}}^{\dagger}_{i}(\bm{k})\hat{{\gamma}}_{j}(\bm{k}) \ket{\Phi}\notag\\
&= A^{2}_{\frac{3}{2}}\bra{\Phi_{\rm{GA}}}\exp\left(\hat{\Pi}_{\frac{3}{2}}\right)\hat{{\gamma}}^{\dagger}_{i}(\bm{k})\hat{{\gamma}}_{j}(\bm{k})\exp\left(\hat{\Pi}_{\frac{3}{2}}^{\dagger}\right)\ket{\Phi_{\rm{GA}}}\notag\\
&= A^{2}_{\frac{3}{2}}\bra{\Phi_{\rm{GA}}}\exp\left(\hat{\Pi}_{\frac{3}{2}}\right)\hat{{\gamma}}^{\dagger}_{i}(\bm{k})\left[\hat{{\gamma}}_{j}(\bm{k}),\exp\left(\hat{\Pi}_{\frac{3}{2}}^{\dagger}\right)\right]\ket{\Phi_{\rm{GA}}}\notag\\
&= \bra{\Phi}\hat{{\gamma}}^{\dagger}_{i}(\bm{k})[\hat{{\gamma}}_{j}(\bm{k}), \hat{\Pi}_{\frac{3}{2}}^{\dagger}]\ket{\Phi}\notag\\
&= \bra{\Phi}\hat{{\gamma}}^{\dagger}_{i}(\bm{k})\left[\frac{1}{2}{\sum_{\bm{k}_{2},\bm{k}_{3}}}'\sum_{a,b=1}^{M}\hat{{\gamma}}^{\dagger}_{a}(\bm{k}_{2})\hat{{\gamma}}^{\dagger}_{b}(\bm{k}_{3}) w_{jab}(\bm{k},\bm{k}_{2},\bm{k}_{3}) \right]\ket{\Phi}\notag\\
&= \frac{1}{2}{\sum_{\bm{k}_{2},\bm{k}_{3}}}'\sum_{a,b=1}^{M}\bra{\Phi}\hat{{\gamma}}^{\dagger}_{i}(\bm{k})\hat{{\gamma}}^{\dagger}_{a}(\bm{k}_{2})\hat{{\gamma}}^{\dagger}_{b}(\bm{k}_{3}) \ket{\Phi} w_{jab}(\bm{k},\bm{k}_{2},\bm{k}_{3})\notag\\
&\simeq\frac{1}{2}{\sum_{\bm{k}_{2},\bm{k}_{3}}}'\sum_{a,b=1}^{M}w^{*}_{iab}(\bm{k},\bm{k}_{2},\bm{k}_{3})w_{jab}(\bm{k},\bm{k}_{2},\bm{k}_{3}).\label{rhoij}
\end{align}}}

\subsection{Ground-state energy and self-consistent conditions}
In this section, we obtain expressions for the ground-state energy and self-consistent equations {embodying} energy-minimum conditions.
We define the following quantities for later convenience:

\begin{subequations}
\begin{align}
\rho_{ij}(\bm{k})&\equiv\bra{\Phi}\hat{d}^{\dagger}_{i}(\bm{k})\hat{d}_{j}(\bm{k})\ket{\Phi},\label{rho}\\
F_{ij}(\bm{k})&\equiv\bra{\Phi}\hat{d}_{i}(\bm{k})\hat{d}_{j}(-\bm{k})\ket{\Phi}\label{F},\\
W_{ij\to j}(\bm{k}_{1},\bm{k}_{2};
-\bm{k}_{3})&\equiv\bra{\Phi}\hat{d}^{\dagger}_{j}(-\bm{k}_{3})\hat{d}_{j}(\bm{k}_{2})\hat{d}_{i}(\bm{k}_{1})\ket{\Phi}\label{W}.
\end{align}
\end{subequations}
Using Eqs.\ (\ref{op1}) and (\ref{op2}), we can transform Eqs.\ (\ref{rho})-(\ref{W}) as follows:

\begin{subequations}
\label{RFW}
\begin{align}
&\rho_{ij}(\bm{k})=\sum_{a=1}^{M}v^{*}_{ia}(\bm{k})v_{aj}(-\bm{k})\notag\\
&+\sum_{a,b=1}^{M}\Big[v^{*}_{ia}(\bm{k}){\rho^{\frac{3}{2}}}^{ *}_{ab}(-\bm{k}) v_{bj}(-\bm{k})
+ u^{*}_{ia}(\bm{k})\rho^{\frac{3}{2}}_{ab}(\bm{k}) u_{bj}(\bm{k})\Big] ,\\
&F_{ij}(\bm{k})=\sum_{a=1}^{M}u_{ia}(\bm{k})v_{aj}(\bm{k})\notag\\
&+\sum_{a,b=1}^{M}\Big[u_{ia}(\bm{k}){\rho^{\frac{3}{2}}}^{ *}_{ab}(\bm{k}) v_{bj}(\bm{k})+ v_{ia}(\bm{k})\rho^{\frac{3}{2}}_{ab}(-\bm{k}) u^{*}_{bj}(-\bm{k})\Big] ,\\
&W_{ij\to j}(\bm{k}_{1},\bm{k}_{2};
-\bm{k}_{3})\notag\\
=&\sum_{a,b,c=1}^{M}\Big[u_{ia}(\bm{k}_{1})u_{jb}(\bm{k}_{2})v^{*}_{jc}(-\bm{k}_{3})w_{abc}(\bm{k}_{1},\bm{k}_{2},\bm{k}_{3})\notag\\
&+u^{*}_{ja}(-\bm{k}_{3})v_{jb}(\bm{k}_{2})v_{ic}(\bm{k}_{1})w^{*}_{abc}(-\bm{k}_{3},-\bm{k}_{2},-\bm{k}_{1})\Big] ,
\end{align}
\end{subequations}
where we have used the symmetries of Eqs.\ (\ref{u}) and (\ref{v}).
We also approximate {{$\hat{c}_{i}(\bm{0})\hat{\beta}^{\dagger}_{i}\simeq \sqrt{N^{\bm{0}}_{i}} e^{i\varphi_{i}}\equiv\psi_{i}$ and $\hat{\beta}_{i}\hat{c}^{\dagger}_{i}(\bm{0})\simeq \sqrt{N^{\bm{0}}_{i}} e^{-i\varphi_{i}}\equiv\psi^{*}_{i}$}}, where $\psi_{i}$ and $\psi^{*}_{i}$ denote homogeneous condensate wave functions of particle $i$, $N^{\bm{0}}_{i}$ denotes the condensed particle number of $i$, and $\varphi_{i}$ denotes the spatially homogeneous phase of condensate wave functions of particle $i$. The total number of particle $i$ is then expressible as
\begin{equation}
N_{i}=N^{\bm{0}}_{i}+{\sum_{\bm{k}}}'\rho_{ii}(\bm{k}).
\label{N_i}
\end{equation}

Now, the ground-state energy can be written {in terms of the quantities in Eqs.\ (\ref{RFW})} and (\ref{N_i}) as follows:
\begin{align}
\mathcal{E}&=\bra{\Phi} \hat{H}\ket{\Phi}\notag\\
&=\sum_{i,j=1}^{M}\frac{U_{ij}}{2\mathcal{V}}N_{i}N_{j}+\sum_{i=1}^{M}{\sum_{\bm{k}}}'\varepsilon^{i}_{k}\rho_{ii}(\bm{k})\notag\\
&+\sum_{i,j=1}^{M}\frac{U_{ij}}{\mathcal{V}}{\sum_{\bm{k}}}'{\rm{Re}}\Big[F^{*}_{ij}({\bm{k}})\psi_{j}\psi_{i} +\rho_{ij}({\bm{k}})\psi_{j}\psi^{*}_{i}\Big]\notag\\
&+\sum_{i,j=1}^{M}\frac{2U_{ij}}{\mathcal{V}}{\sum_{\bm{k}_{1},\bm{k}_{2},\bm{k}_{3}}}^{'}\delta_{\bm{k}_{1}+\bm{k}_{2}+\bm{k}_{3},\bm{0}}{\rm{Re}}\left[\psi^{*}_{i}W_{ij\to j}(\bm{k}_{1},\bm{k}_{2};
-\bm{k}_{3})\right] \notag\\
&+\sum_{i,j=1}^{M}\frac{U_{ij}}{2\mathcal{V}}{\sum_{\bm{k},\bm{k}'}}^{'}\Big[F_{ij}(\bm{k})F^{*}_{ij}(\bm{k}')+\rho_{ij}(\bm{k})\rho^{*}_{ij}(\bm{k}')\Big],
\label{calE}
\end{align}
{{where we use the Wick decomposition 
\begin{align}
&\bra{\Phi}\hat{d}^{\dagger}_{i}(\bm{k}+\bm{q})\hat{d}^{\dagger}_{j}(\bm{k}'-\bm{q})\hat{d}_{j}(\bm{k}')\hat{d}_{i}(\bm{k}) \ket{\Phi}\notag\\
&\simeq \delta_{\bm{q},\bm{0}}\rho_{ii}(\bm{k})\rho_{jj}(\bm{k}')+{{\delta_{\bm{k}',\bm{k}+\bm{q}}\rho^{*}_{ij}(\bm{k})\rho_{ij}(\bm{k}')}}\notag\\
&+\delta_{\bm{k}',-\bm{k}}F_{ij}(\bm{k})F^{*}_{ij}(\bm{k}+\bm{q}).
\end{align}}}

We now minimize Eq.\ (\ref{calE}) under the constraints of Eq.\ (\ref{N_i}).
This can be performed most easily in terms of $\Omega\equiv \mathcal{E}-\sum_{i}\mu_{i}N_{i}$, where $\mu_i$ denotes the Lagrange multipliers. Specifically, we
 determine $\psi_{i}$, $\phi_{ij}(\bm{k})$, and $w_{ijk}(\bm{k}_{1},\bm{k}_{2},\bm{k}_{3})$ from the stationarity conditions
\begin{equation}
\frac{\delta \Omega}{\delta \psi^{*}_{i}}=0, \ \frac{\delta \Omega}{\delta \phi^{*}_{ij}(\bm{k})}=0, \ \frac{\delta \Omega}{\delta w^{*}_{ijk}(\bm{k}_{1},\bm{k}_{2},\bm{k}_{3})}=0. \label{self}
\end{equation}
When considering the system composed of $M\geq 3$ types of bosons, we need to solve Eq.\ (\ref{self}) simultaneously with $M+M^{2}+M^{3}$ types of variational functions in principle. 

Finally, we comment about the phases of the variational parameters.
From our numerical calculations for $M=2$ systems, all the variational parameters turned out to be real numbers.
However, the variational parameters for $M\geq3$ systems may have phases, meaning that Eq.\ (\ref{self}) should be calculated with their imaginary parts in general.

\subsection{Correction to the stability condition for $2$-component systems}
If $U_{ij}$ $(i\neq j)$ is too strong in a homogeneous $2$-component system, it is known that the system becomes unstable by (i) forming denser states containing both components called droplets\cite{Th6} when $U_{i j}<0$, or (ii) causing a phase separation into two components\cite{Th6pl} when $U_{ij}>0$.
The stability condition for a homogeneous system is given by $U^{2}_{ij}<U_{ii}U_{jj}$\cite{Pethic}, which is derived by neglecting $3/2$-body and $2$-body processes. 
In this section, we reconsider the stability condition for a $2$-component system composed of particles $A$ and $B$ on the basis of the ground-state wave function including $3/2$-body and $2$-body processes.
Here, we assume that all the variational parameters are real numbers.
Under this assumption, the functional $\Omega$ is given in terms of variational parameters by
\begin{equation}
\Omega=\Omega[\psi_{A},\psi_{B},\phi_{AA},\phi_{AB}=\phi_{BA},\phi_{BB},w_{AAA},w_{ABA},w_{BAB},w_{BBB}].
\end{equation}
For the homogeneous solution to be stable, $\Omega$ must have a minimum value with respect to all the variational parameters and the second-order variation of $\Omega$ must always be positive. Therefore,
\begin{align}
\delta^{2}\Omega=\bm{\eta}^{T}A\bm{\eta}>0,
\end{align}
where $\bm{\eta}$ is a column vector composed of small variations in all the variational parameters the entire $\bm{k}$ space and $A$ is the corresponding Hessian matrix\cite{Hessian}. 

To consider the complete condition that $\Omega$ has a minimum value, all the eigenvalues of $A$ must be positive, i.e., ${\rm{det}}A>0$.
However, it is difficult to show this completely both analytically and computationally because $A$ is quite a large matrix.
Here, we consider some necessary conditions for $\Omega$ to have minimum value,
\begin{align}
\frac{\partial^{2} \Omega}{\partial \psi^{2}_{i}}> 0 , \ \ {\rm{det}}A_{\psi_{A}\psi_{B}}> 0 , \label{detsmall}
\end{align}
where $A_{\psi_{A}\psi_{B}}$ is a submatrix of $A$ defined by
\begin{align}
A_{\psi_{A}\psi_{B}}\equiv \begin{pmatrix}
\dfrac{\partial^{2} \Omega}{\partial \psi^{2}_{A}}& \dfrac{\partial^{2} \Omega}{\partial \psi_{A}\partial \psi_{B}} \\
\dfrac{\partial^{2} \Omega}{\partial \psi_{B}\partial \psi_{A}}& \dfrac{\partial^{2} \Omega}{\partial \psi^{2}_{B}}
\end{pmatrix}.
\end{align}
$\partial^{2} \Omega/\partial \psi_{i}\partial \psi_{j}$ is calculated by
\begin{align}
\frac{\partial^{2} \Omega}{\partial \psi_{i}\partial \psi_{j}}=\frac{4\sqrt{N^{\bm{0}}_{i}N^{\bm{0}}_{j}}}{\mathcal{V}}U_{ij}(1+c_{ij}),\label{Uij} \ 
\end{align}
where
\begin{subequations}
\begin{align}
c_{ii} =&
\frac{-N_{\rm{all}}}{2U_{ii}N^{\bm{0}}_{i}} \Bigg [\sum_{i'=A,B}(1-\delta_{i,i'})\sqrt{\frac{N^{\bm{0}}_{i'}}{N^{\bm{0}}_{i}}} \frac{U_{AB}}{N_{\rm{all}}}{\sum_{\bm{k}}}'\Big\{F_{AB}({\bm{k}}) +\rho_{AB}({\bm{k}})\Big\}\notag\\
&+\frac{N_{\rm{all}}}{\sqrt{{N^{\bm{0}}}_{i}}}\sum_{i'=A,B}\frac{U_{ii'}}{N^{2}_{\rm{all}}}{\sum_{\bm{k}_{1},\bm{k}_{2},\bm{k}_{3}}}^{'}{\delta_{\bm{k}_{1}+\bm{k}_{2}+\bm{k}_{3}\bm{0}}} \Big\{W_{ii'\to i'}(\bm{k}_{1},\bm{k}_{2};
-\bm{k}_{3})\Big\} \Bigg],\label{coeff1}\\
c_{AB}& =\frac{N_{\rm{all}}}{2\sqrt{N^{\bm{0}}_{A}N^{\bm{0}}_{B}}}\frac{1}{N_{\rm{all}}}{\sum_{\bm{k}}}'\Big[F_{AB}({\bm{k}}) +\rho_{AB}({\bm{k}}) \Big]=c_{BA}.\label{coeff2}
\end{align}
\end{subequations}
The first condition of Eq.\ (\ref{detsmall}) with Eqs.\ (\ref{Uij}), (\ref{coeff1}), and (\ref{coeff2}) requires the relation $U_{ii}>0$ since $1+c_{ii}$ with $|c_{ii}|\ll1$ is always positive.
On the other hand, the second condition of Eqs.\ (\ref{detsmall}) with (\ref{Uij}), (\ref{coeff1}) and (\ref{coeff2}) gives the following stability condition:
\begin{align}
&U_{AA}U_{BB}\Big(1+c_{AA}\Big)\Big(1+c_{BB}\Big)-U^{2}_{AB}\Big(1+c_{AB}\Big)^{2} > 0\notag\\
\to&\frac{U^{2}_{AB}}{U_{AA}U_{BB}}< \frac{1+(c_{AA}+c_{BB})+c_{AA}c_{BB}}{1+2c_{AB}+c^{2}_{AB}}\equiv 1+\alpha,\label{stab}
\end{align}
where $\alpha$ is the correction value, which is determined after solving Eq.\ (\ref{self}) self-consistently and obtaining $\ket{\Phi}$.
The conventional relation $U^{2}_{AB}/U_{AA}U_{BB}<1$ is obtained by putting $\alpha=0$ ($c_{AA}=c_{AB}=c_{BB}=0$), which corresponds to the calculation with $w_{AAA},w_{BBB},w_{ABA},w_{BAB}\to 0$ and $N^{\bm{0}}_{i}\to N_{i}$. 
In the following, we show $\alpha\neq 0$ numerically and $2$-body and $3/2$-body interactions are considered.

\section{Numerical Calculation}
In this section, we consider a system composed of two kinds of species labeled by $A$ and $B$ and calculate Eq.\ (\ref{self}) self-consistently with respect to variational parameters $\psi_{A}$, $\psi_{B}$, $\phi_{AA}$, $\phi_{AB}$, $\phi_{BB}$, $w_{AAA}$, $w_{ABA}$, $w_{BAB}$ and $w_{BBB}$.
We outline the numerical procedures and show the results of (i) ground-state energies, (ii) variational parameters {{$\phi_{AA}(k)$, $\phi_{BB}(k)$ and $\phi_{AB}(k)$}}, {{and (iii) $c_{AA}$, $c_{AB}$, $c_{BB}$, and $\alpha$}}. Our numerical procedures mentioned below reduce to the ones given in Sec.\ 3 in Ref.\ \citen{kita3/2} when $N_{A}=N$ and $N_{B}=0$.
\subsection{Numerical procedures}
{First, we introduce the expression for the effective interaction potential between particles $i$ and $j$ as follows:\cite{Pethic}
\begin{equation}
U_{ij}=\frac{2\pi\hbar^{2} a_{U_{ij}}}{m_{ij}},
\end{equation}
where $m_{ij}\equiv m_{i}m_{j}/(m_{i}+m_{j})$.
The ultraviolet divergence inherent in the potential is removed by introducing a cutoff wavenumber $k_{\rm{c}}$ into every summation over $k$ as
\begin{equation}
{\sum_{\bm{k}}}^{'}\to {\sum_{\bm{k}}}^{'}\theta(k_{\rm{c}}-k).
\end{equation}
Similarly to the single-component systems, the $s$-wave scattering length $a_{AA}$, which originates from $U_{AA}$ is obtained by
\begin{equation}
\frac{m_{i}}{4\pi \hbar^{2} a_{ii}}=\frac{1}{U_{ii}}+\int \frac{d^{3}k}{(2\pi)^{3}} \frac{\theta(k_{\rm{c}}-k)}{2\varepsilon^{i}_{k}},
\end{equation}
which yields
\begin{equation}
a_{ii}=\frac{a_{U_{ii}}}{1+2k_{\rm{c}}a_{U_{ii}}/\pi}.
\end{equation}
In the following calculations, we choose $k_{\rm{c}}$ that satisfies $k_{\rm{c}}a_{U_{ii}}\ll 1$ (i.e., $\ a_{ii}\simeq a_{U_{ii}})$.
}

Next, we introduce the units of energy and wavenumber for performing the numerical calculations.
The characteristic energy and wavenumber of this system are defined by
\begin{equation}
\varepsilon_{U_{AA}}\equiv \bar{n}_{\rm{all}}U_{AA},\ k_{U_{AA}}=\sqrt{8\pi a_{U_{AA}}\bar{n}_{\rm{all}}},
\end{equation}
where $\bar{n}_{\rm{all}}\equiv N_{\rm{all}}/\mathcal{V}\equiv \bar{n}_{A}+\bar{n}_{B}$.

Hereafter, we use the following dimensionless coupling constants;
\begin{equation}
{{\delta_{A}\equiv a^{3}_{U_{AA}}\bar{n}_{\rm{all}}, \ \delta_{B}\equiv a^{3}_{U_{BB}}\bar{n}_{\rm{all}}, \delta_{AB}\equiv a^{3}_{U_{AB}}\bar{n}_{\rm{all}}.}}
\end{equation}
{{Using these parameters,}} $U_{BB}/U_{AA}$ and $U_{AB}/U_{AA}$ are given by
\begin{equation}
\frac{U_{BB}}{U_{AA}}=\frac{\delta^{\frac{1}{3}}_{B} m_{A}}{\delta^{\frac{1}{3}}_{A}m_{B}}, \  \frac{U_{AB}}{U_{AA}}=\frac{1}{2}\frac{\delta^{\frac{1}{3}}_{AB} }{\delta^{\frac{1}{3}}_{A}}\left(1+\frac{m_{A}}{m_{B}}\right).\label{delta}
\end{equation} 
{{$\delta_{AB}$ can be obtained from Eq.\ (\ref{delta}) since we set $\delta_{A}$, $\delta_{B}$, $m_{A}/m_{B}$ and $U_{AB}/\sqrt{U_{AA}U_{BB}}$ as external parameters.}}
Sums over $\bm{k}$ are transformed into integrals as follows \cite{kita3/2}:

\begin{subequations}
\begin{align}
&\frac{1}{N_{\rm{all}}}{\sum_{\bm{k}}}' =\sqrt{\frac{128\delta_{A}}{\pi}}\int _{0}^{\tilde{k}_{\rm{c}}}d\tilde{k}\tilde{k}^{2}\\
&\frac{1}{N_{\rm{all}}}{\sum_{\bm{k}_{2},\bm{k}_{3}}}' \delta_{\bm{k}+\bm{k}_{2}+\bm{k}_{3},\bm{0}}\notag\\
&=\frac{1}{2\tilde{k}}\sqrt{\frac{128\delta_{A}}{\pi}}\int _{0}^{\tilde{k}_{\rm{c}}}d\tilde{k}_{2}\tilde{k}_{2}\int _{|\tilde{k}-\tilde{k}_{2}|}^{{\rm{min}}(\tilde{k}+\tilde{k}_{2},\tilde{k}_{\rm{c}})}d\tilde{k}_{3}\tilde{k}_{3},
\end{align}
\end{subequations}
where $\tilde{k}\equiv k/k_{U_{AA}}$.

To carry out numerical calculations, we need to obtain the analytic expressions for $u_{ij}(k)$, $v_{ij}(k)$, $\sum_{i'j'}\sum'_{\bm{k}'}\delta {u_{i'j'}(k')}/\delta\phi_{ij}(k)$, and $\sum_{i'j'}\sum_{\bm{k}'}\delta {v_{i'j'}(k')}/\delta\phi_{ij}(k)$.
Considering that $\Phi_{ij}(k)$ is a Hermitian matrix, $u_{ij}(k)$ is obtained by 
\begin{align}
 \underline{u}=\underline{\Phi}^{-\frac{1}{2}}=(\underline{P}^{-1}\underline{\Lambda}^{\frac{1}{2}} \underline{P})^{-1}\label{sqrt},
\end{align}
where $\underline{P}$ denotes a $2\times 2$ unitary matrix that diagonalizes $\underline{\Phi}$ and $\Lambda\equiv {\rm{diag}}(\lambda_{1},\lambda_{2})$ is the diagonal matrix with eigenvalues $\lambda_{1}$ and $\lambda_{2}$.
From Eq.\ (\ref{sqrt}), we can obtain the analytic forms of $u_{ij}(k)$, $v_{ij}(k)$, $\sum_{i'j'}\sum'_{\bm{k}'}\delta {u_{i'j'}(k')}/\delta\phi_{ij}(k)$, and {{$\sum_{i'j'}\sum'_{\bm{k}'}\delta {v_{i'j'}(k')}/\delta\phi_{ij}(k)$}}.
However, because of the huge number of terms, it is difficult to perform this calculation by hand.
To deal with this problem, we used Mathematica and obtained analytic expressions.

Finally, we sketch the numerical procedures.
We started the initial self-consistent calculation by substituting the trivial solutions for $U_{AB}=0$ given by Ref. \citen{kita3/2} and renewed the solutions one after another.
To avoid the irregular numerical fluctuation of variational parameters, self-consistent calculation was carefully performed by mixing the old and new solutions with weight ratio $95:5$.
The convergence of the iteration can be checked by monitoring the ground-state energy.
We stopped the iteration when the magnitude of the relative difference between the old and new energies decreased to below $10^{-10}$.

\subsection{{Ground-state energies, variational parameters, and corrections to the stability condition}}

\begin{table}[t]
\caption{{{{$\tilde{\mathcal{E}}^{(0)}_{\rm{eff}}$, $\tilde{\mathcal{E}}^{(1)}_{\rm{eff}}$, and $\Delta\tilde{\mathcal{E}}$ in various cases and conditions with {{$\tilde{k}_{\rm{c}}=5$} and} $|U_{AB}|/\sqrt{U_{AA}U_{BB}}=0.95$.}}}}
\centering
\begin{tabular}{c c c c c c c c}
\hline\hline\\[-2.0ex]
{{Case}} & ${{\tilde{\mathcal{E}}^{(0)}_{\rm{eff}}}}$  & ${{\tilde{\mathcal{E}}^{(1)}_{\rm{eff}}}}$ & ${{\Delta\tilde{\mathcal{E}}_{(I)}}}$& ${{\Delta\tilde{\mathcal{E}}_{(I\hspace{-.1em}I)}}}$ & ${{\Delta\tilde{\mathcal{E}}_{(I\hspace{-.1em}I\hspace{-.1em}I)}}}$ & ${{\Delta\tilde{\mathcal{E}}_{(I\hspace{-.1em}V)}}}$& ${{\Delta\tilde{\mathcal{E}}_{(V)}}}$ \\ [0.5ex] % inserts table %heading
\hline
{{($A_{+}$)}}&$ {{0.488}} $&$ {{-5.62}} $&$ {{2730}} $&$ {{80.9}} $ & $ {{79.0}}$&$ {{62.4}}$&$ {{22.8}} $ \\[0.2ex] 
{{($B_{+}$)}}&$ {{0.492}} $&$ {{-5.72}} $&$ {{1530}} $&$ {{82.9}} $ & $ {{81.0}}$&$ {{49.3}}$&$ {{24.2}}$ \\[0.2ex] 
{{($C_{+}$)}}&$ {{1.10}} $&$ {{-11.9}} $&$ {{3800}} $&{{162}} $ $ & $ {{158}}$&$ {{109}}$&$ {{52.7}}$\\[0.2ex]  \hline
{{($A_{-}$)}}&$ {{0.0125}} $&$ {{-5.62}} $&$ {{2730}} $&$ {{18.2}} $& $ {{16.9}}$ &$ {{1.29}} $&$ {{-3.69}}$ \\[0.2ex] 
{{($B_{-}$)}}&$ {{0.188}} $&$ {{-5.72}} $&$ {{1530}} $&$ {{42.7}} $& $ {{41.2}}$ &$ {{10.2}} $&$ {{7.01}}$ \\[0.2ex] 
{{($C_{-}$)}}&$ {{0.150}} $&$ {{-11.9}} $&$ {{3800}} $&$ {{81.9}}  $ & $ {{79.0}}$&$ {{31.3}} $&$ {{18.5}} $\\[0.2ex] 
\hline
\end{tabular}
\label{GE1}
\end{table}

First, we estimate the ground-state energy.
We show that the ground state incorporating $O(\sqrt{N}_{i})$ terms of the Hamiltonians in Eqs.\ (\ref{3/2body}) (3/2-body interaction) and (\ref{2body}) (2-body interaction) gives lower energy than the one given by the eigenstate of the approximated Hamiltonian $\hat{H}_{\rm{Bog}}\equiv \hat{H}_{0}+\hat{H}_{1}$\cite{Fetter}.
To see this clearly, we diagonalize $\hat{H}_{\rm{Bog}}$ and obtain the ground-state energy as follows:
\begin{equation}
\mathcal{E}_{\rm{eff}}= \mathcal{E}^{(0)}_{\rm{eff}}+\mathcal{E}^{(1)}_{\rm{eff}},
\end{equation}
where we define the following quantities:
 \begin{subequations}
\begin{align}
&
\mathcal{E}^{(0)}_{\rm{eff}}\equiv \sum_{i=A,B}\sum_{j=A,B}\frac{U_{ij}}{2\mathcal{V}}N_{i}N_{j},\\
&{{\mathcal{E}^{(1)}_{\rm{eff}}\equiv -\frac{1}{2}{\sum_{\bm{k}}}^{'}\Big[\sum_{i=A,B}E_{i}(k)-\sum_{\sigma=+,-}E_{\sigma}(k)\Big].}}\\
&E_{i}(k)\equiv \varepsilon^{i}_{k}+\bar{n}_{i}U_{ii},\\
&E_{\pm}(k)\equiv \frac{1}{\sqrt{2}}\Bigg[\big(E^{\rm{Bog}}_{A}(k))^{2}+\big(E^{\rm{Bog}}_{B}(k))^{2}\notag\\
&\pm\sqrt{\Big\{\big(E^{\rm{Bog}}_{A}(k))^{2}-\big(E^{\rm{Bog}}_{B}(k))^{2}\Big\}^{2}+16\bar{n}_{A}\bar{n}_{B} \varepsilon^{A}_{k}\varepsilon^{B}_{k}U^{2}_{AB}}\Bigg]^{\frac{1}{2}}\label{mixedBog},\\
&E^{\rm{Bog}}_{i}(k)\equiv \sqrt{\varepsilon^{i}_{k}(\varepsilon^{i}_{k}+2\bar{n}_{i}U_{ii})}.
\end{align}
 \end{subequations}
We have confirmed that {{$\bra{\Phi}\hat{H}_{\rm{Bog}}\ket{\Phi}$}} estimated by our variational calculations with {$w_{AAA},w_{BBB},w_{ABA},w_{BAB}\to 0$} and $N^{\bm{0}}_{i}\to N_{i}$ coincides with $\mathcal{E}_{\rm{eff}}$ numerically.

\begin{table}[t]
\caption{{{{$\tilde{\mathcal{E}}^{(0)}_{\rm{eff}}$, $\tilde{\mathcal{E}}^{(1)}_{\rm{eff}}$, and $\Delta\tilde{\mathcal{E}}$ in various cases and conditions with {{$\tilde{k}_{\rm{c}}=10$} and} $|U_{AB}|/\sqrt{U_{AA}U_{BB}}=0.95$.}}}}
\centering
\begin{tabular}{c c c c c c c c}
\hline\hline\\[-2.0ex]
Case & ${{\tilde{\mathcal{E}}^{(0)}_{\rm{eff}}}}$  & ${{\tilde{\mathcal{E}}^{(1)}_{\rm{eff}}}}$ & $\Delta\tilde{\mathcal{E}}_{(I)}$& $\Delta\tilde{\mathcal{E}}_{(I\hspace{-.1em}I)}$ & $\Delta\tilde{\mathcal{E}}_{(I\hspace{-.1em}I\hspace{-.1em}I)}$ & $\Delta\tilde{\mathcal{E}}_{(I\hspace{-.1em}V)}$& $\Delta\tilde{\mathcal{E}}_{(V)}$ \\ [0.5ex] % inserts table %heading
\hline
($A_{+}$)&${{0.488}}$&${{-13.0}}$&$ 6120 $&$ 392 $ &$ 377$& $ 335 $&$ 230 $ \\[0.2ex] 
($B_{+}$)&${{0.492}}$&${{-13.3}}$&$ 4080 $&$ 402  $ &$ 387 $& $ 304 $&$ 237 $ \\[0.2ex] 
($C_{+}$)&${{1.10}}$&${{-27.4}}$&$ 9990 $&$ 770 $ &$ 743 $& $ 623 $&$ 472 $\\[0.2ex]  \hline
($A_{-}$)&${{0.0125}}$&${{-13.0}}$&$ 6120 $&$ 62.6 $ &$ 50.5 $& $10.3 $&$ {{-5.24}}$ \\[0.2ex] 
($B_{-}$)&${{0.188}}$&${{-13.3}}$&$ 4080 $&$ 191  $ &$ 178 $& $ 96.3 $&$86.1$ \\[0.2ex] 
($C_{-}$)&${{0.150}}$&${{-27.4}}$&$ 9990 $&$ 350  $ &$ 326 $& $ 208 $&$ 171 $\\[0.2ex] 
\hline
\end{tabular}
\label{GE2}
\end{table}

\begin{table}[t]
\caption{{{{$\tilde{\mathcal{E}}^{(0)}_{\rm{eff}}$, $\tilde{\mathcal{E}}^{(1)}_{\rm{eff}}$, and $\Delta\tilde{\mathcal{E}}$ in various cases and conditions with {{$\tilde{k}_{\rm{c}}=5$} and} $|U_{AB}|/\sqrt{U_{AA}U_{BB}}=0.98$.}}}}
\centering
\begin{tabular}{c c c c c c c c}
\hline\hline\\[-2.0ex]
{{Case}} & ${{\tilde{\mathcal{E}}^{(0)}_{\rm{eff}}}}$  & ${{\tilde{\mathcal{E}}^{(1)}_{\rm{eff}}}}$ & ${{\Delta\tilde{\mathcal{E}}_{(I)}}}$& ${{\Delta\tilde{\mathcal{E}}_{(I\hspace{-.1em}I)}}}$ & ${{\Delta\tilde{\mathcal{E}}_{(I\hspace{-.1em}I\hspace{-.1em}I)}}}$ & ${{\Delta\tilde{\mathcal{E}}_{(I\hspace{-.1em}V)}}}$& ${{\Delta\tilde{\mathcal{E}}_{(V)}}}$ \\ [0.5ex] % inserts table %heading
\hline
{{($A_{+}$)}}&${{ 0.495 }} $&$ {{ -5.78 }} $&$ {{ 2610 }} $&$ {{84.2}} $ &$ {{82.2}} $&$ {{ 66.2 }} $&$ {{ 23.3 }} $\\[0.2ex] 
{{($B_{+}$)}}&${{ 0.497 }}$&$ {{ -5.82 }} $&$ {{ 1710 }} $&$ {{85.0}} $ &$ {{83.1}} $&$ {{51.8}} $&$ {{ 24.5 }} $\\[0.2ex] 
{{($C_{+}$)}}&$ {{ 1.12 }} $&$ {{ -12.1 }} $&$ {{ 4210 }} $&$ {{166}} $ &$ {{162}} $&$ {{114}} $&$ {{ 53.0 }} $\\[0.2ex]  \hline
{{($A_{-}$)}}&$ {{ 0.00500 }} $&$ {{ -5.78 }} $&$ {{ 2610 }} $&$ {{15.2}} $ &$ {{13.8}} $&$ {{-1.19}} $&$ {{ -5.87 }} $ \\[0.2ex] 
{{($B_{-}$)}}&$ {{ 0.183 }} $&$ {{ -5.82 }} $&$ {{ 1710 }} $&$ {{40.8}} $ &$ {{39.3}} $&$ {{8.72}} $&$ {{ 5.62 }} $\\[0.2ex] 
{{($C_{-}$)}}&$ {{ 0.135 }} $&$ {{ -12.1 }} $&$ {{ 4210 }} $&$ {{77.9}} $ &$ {{75.0}} $&$ {{28.2}} $&$ {{ 15.3}} $\\[0.2ex] 
\hline
\end{tabular}
\label{GE3}
\end{table}

Since our interest is to estimate the ground-state energies including $\hat{H}_{3/2}$ and $\hat{H}_{2}$, we calculate the quantities defined by 
 {{
 \begin{subequations}
 \begin{align}
&\tilde{\mathcal{E}}^{(0)}_{\rm{eff}}\equiv \mathcal{E}^{(0)}_{\rm{eff}}/(N_{\rm{all}}\varepsilon_{U_{AA}})=\frac{1}{2}\sum_{i=A,B}\sum_{j=A,B}\frac{N_{i}N_{j}}{N^{2}_{\rm{all}}}\frac{U_{ij}}{U_{AA}}, \\
&\tilde{\mathcal{E}}^{(1)}_{\rm{eff}}\equiv \mathcal{E}^{(1)}_{\rm{eff}}/(N_{\rm{all}}\varepsilon_{U_{AA}} )\times \delta^{-\frac{1}{2}}_{A},\\
&\Delta\tilde{\mathcal{E}}\equiv (\mathcal{E}-\mathcal{E}_{\rm{eff}})/(N_{\rm{all}}\varepsilon_{U_{AA}})\times \delta^{-1}_{A},
  \end{align}
 \end{subequations}
 }
 } 
 and evaluate their values for the six cases
\begin{itemize}
\item[($A_{\pm}$)]$m_{A}:m_{B}=\bar{n}_{A}:\bar{n}_{B}=1:1$,
\item[($B_{\pm}$)] $m_{A}:m_{B}=1:1$, $\bar{n}_{A}:\bar{n}_{B}=1:4$,
\item[($C_{\pm}$)] $m_{A}:m_{B}=4:1$, $\bar{n}_{A}:\bar{n}_{B}=1:1$,
\end{itemize}
where $\pm$ denotes the sign of $U_{AB}$.
We set the other parameters as $\delta_{A}=\delta_{B}=1.0\times 10^{-6}$ and $N_{\rm{all}}=10^{8}$.

 Incorporating more variational parameters in the theory is expected to yield a better estimate for the ground-state energy.
To see this explicitly, we have performed our variational calculations for the following five conditions.
\begin{itemize}
\item[$(I)$] $N^{\bm{0}}_{i}\to N_{i}$ and {$\phi_{AB}=w_{AAA}=w_{BBB}=w_{ABA}=w_{BAB}=0$}. This case corresponds to the Bogoliubov theory with no correlations between different species.
\item[$(I\hspace{-.1em}I)$] $N^{\bm{0}}_{i}\to N_{i}$ and {$w_{AAA}=w_{BBB}=w_{ABA}=w_{BAB}=0$}. This case corresponds to the eigenstate of $\hat{H}_{\rm{Bog}}$ or standard multi-component Gross--Pitaevskii theory \cite{Pethic}.
\item[$(I\hspace{-.1em}I\hspace{-.1em}I)$] {$w_{AAA}=w_{BBB}=w_{ABA}=w_{BAB}=0$}. This case corresponds to $\ket{\Phi_{\rm{GA}}}$.
\item[$(I\hspace{-.1em}V)$] {$w_{ABA}=w_{BAB}=0$}.
\item[$(V)$] All the variational parameters are calculated self-consistently.
\end{itemize}
The corresponding energies are denoted by $\Delta\tilde{\mathcal{E}}_{(I)}$, $\Delta\tilde{\mathcal{E}}_{(I\hspace{-.1em}I)}$, $\Delta\tilde{\mathcal{E}}_{(I\hspace{-.1em}I\hspace{-.1em}I)}$, $\Delta\tilde{\mathcal{E}}_{(I\hspace{-.1em}V)}$, and $\Delta\tilde{\mathcal{E}}_{(V)}$.
As shown in Tables \ref{GE1} - \ref{GE3}, we can confirm the relation $\Delta\tilde{\mathcal{E}}_{(I)}\gg\Delta\tilde{\mathcal{E}}_{(I\hspace{-.1em}I)}>\Delta\tilde{\mathcal{E}}_{(I\hspace{-.1em}I\hspace{-.1em}I)}>\Delta\tilde{\mathcal{E}}_{(I\hspace{-.1em}V)}>\Delta\tilde{\mathcal{E}}_{(V)}$ for all the cases of ($A_{\pm}$), ($B_{\pm}$), and ($C_{\pm}$).
Therefore, the ground state of a $2$-component miscible BEC with the contributions from $2$-body and $3/2$-body processes is constructed through these self-consistent calculations.
In addition, as we see from the tables, $|\Delta\tilde{\mathcal{E}}_{(I\hspace{-.1em}I)}-\Delta\tilde{\mathcal{E}}_{(I\hspace{-.1em}I\hspace{-.1em}I)}|<|\Delta\tilde{\mathcal{E}}_{(I\hspace{-.1em}I\hspace{-.1em}I)}-\Delta\tilde{\mathcal{E}}_{(V)}|$ in all the cases.
This result indicates that $3/2$-body processes contribute to lowering the ground-state energies more than $2$-body processes.
In this sense, the mean-field approximation for mixed BECs is not quantitatively effective even in the weak-coupling region, as well as the single-component systems\cite{kita3/2}.

\begin{figure}[t]
        \begin{center}
                \includegraphics[width=0.6\linewidth]{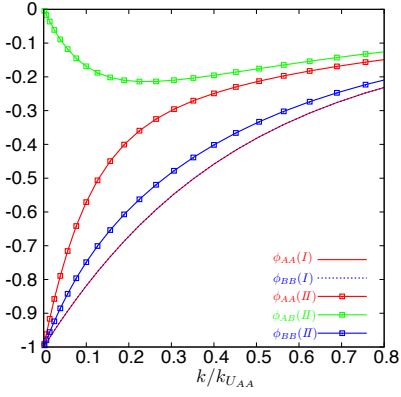}
                \end{center}
\caption{
(Color online)  { $\phi_{AA}(k)$ (red lines), $\phi_{AB}(k)$ (green line), and $\phi_{BB}(k)$ (blue lines) given by the case $m_{A}:m_{B}=4:1$, $n_{A}:n_{B}=1:1$, $\tilde{k}_{\rm{c}}=10$, and $U_{AB}=0.95\sqrt{U_{AA}U_{BB}}$ with conditions $(I)$ (solid and short dashed lines) and $(I\hspace{-.1em}I)$ (square-point lines)}. 
}
\label{fig1}
\end{figure}

We also obtain the variational parameters $\phi_{AA}(k)$, $\phi_{BB}(k)$, and $\phi_{AB}(k)$ which characterize the pair excitations of particles with wave numbers $\bm{k}$ and $-\bm{k}$ from condensates.
Figure \ref{fig1} shows the behavior of $\phi_{AA}$ (red lines), $\phi_{BB}$ (blue lines), and $\phi_{AB}$ (green lines) for cases ($C_{\pm}$) with conditions $(I)$ (solid and short dashed lines) and $(I\hspace{-.1em}I)$ (square-point lines).
The variational parameters for $(I)$ are analytically given by
\begin{align}
\phi_{ij}(k)=-\delta_{i,j}\frac{\varepsilon^{i}_{k}+\bar{n}_{i}U_{ii}-E^{\rm{Bog}}_{i}(k)}{\bar{n}_{i}U_{ii}}.\label{phiij}
\end{align}
Specifically, when $a_{U_{AA}}=a_{U_{BB}}$ and $\bar{n}_{A}=\bar{n}_{B}$, such as in cases $(A_{\pm})$ or $(C_{\pm})$, we find that $\phi_{AA}=\phi_{BB}$ regardless of the mass parameters, as shown in Fig.\ \ref{fig1}, which originates from the relation $U_{AA}/U_{BB}=m_{ B}/m_{A}$.
On the other hand, the behaviors of $\phi_{AA}$ and $\phi_{BB}$ for $(I\hspace{-.1em}I)$ are clearly different from each other.
Therefore, we see that pair excitations between different particles characterize the mass difference between them.
Furthermore, we have numerically checked that the term proportional to ${{U_{AB}\sum'_{\bm{k}}[\rho_{AB}(k)+F_{AB}(k)]}}$ in the ground-state energy is always negative and of order $\delta^{\frac{1}{2}}_{A}$ .
Thus, we find that pair excitation between different particles dramatically lowers the ground-state energy as shown in Tables \ref{GE1} - \ref{GE3}.

\begin{figure}[t]
        \begin{center}
                \includegraphics[width=0.6\linewidth]{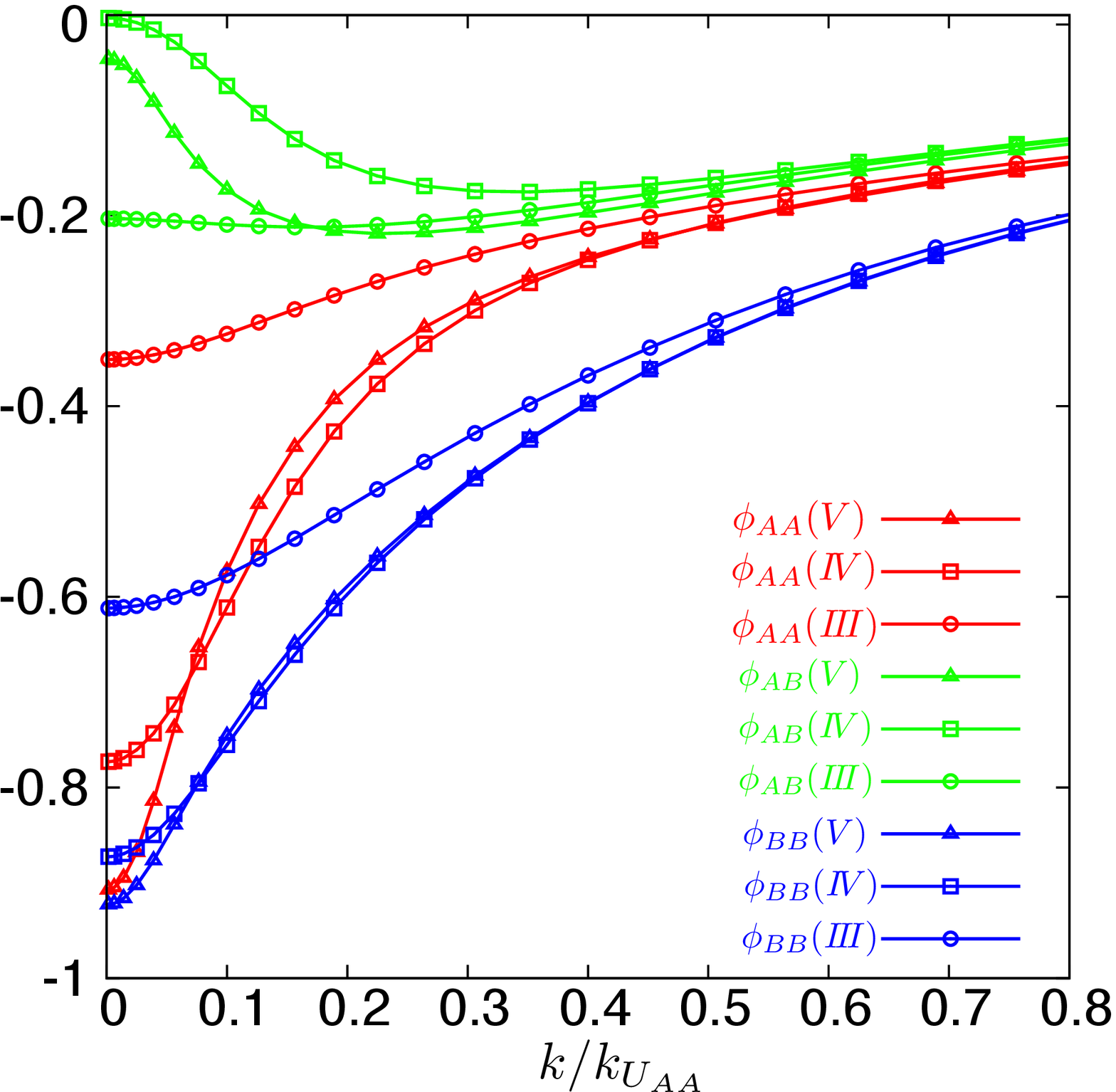}
                \end{center}
\caption{
(Color online)  Variational parameters  $\phi_{AA}(k)$ {(red lines)}, $\phi_{AB}(k)$ (green lines), and $\phi_{BB}(k)$ (blue lines) given by the case $m_{A}:m_{B}=4:1$, $n_{A}:n_{B}=1:1$, $\tilde{k}_{\rm{c}}=10$, and $U_{AB}=0.95\sqrt{U_{AA}U_{BB}}$ with conditions $(I\hspace{-.1em}I\hspace{-.1em}I)$ (circular-point lines), $(I\hspace{-.1em}V)$ (square-point lines), and $(V)$ (triangular-point lines). 
}
\label{fig2}
\end{figure}
\begin{figure}
        \begin{center}
                \includegraphics[width=0.6\linewidth]{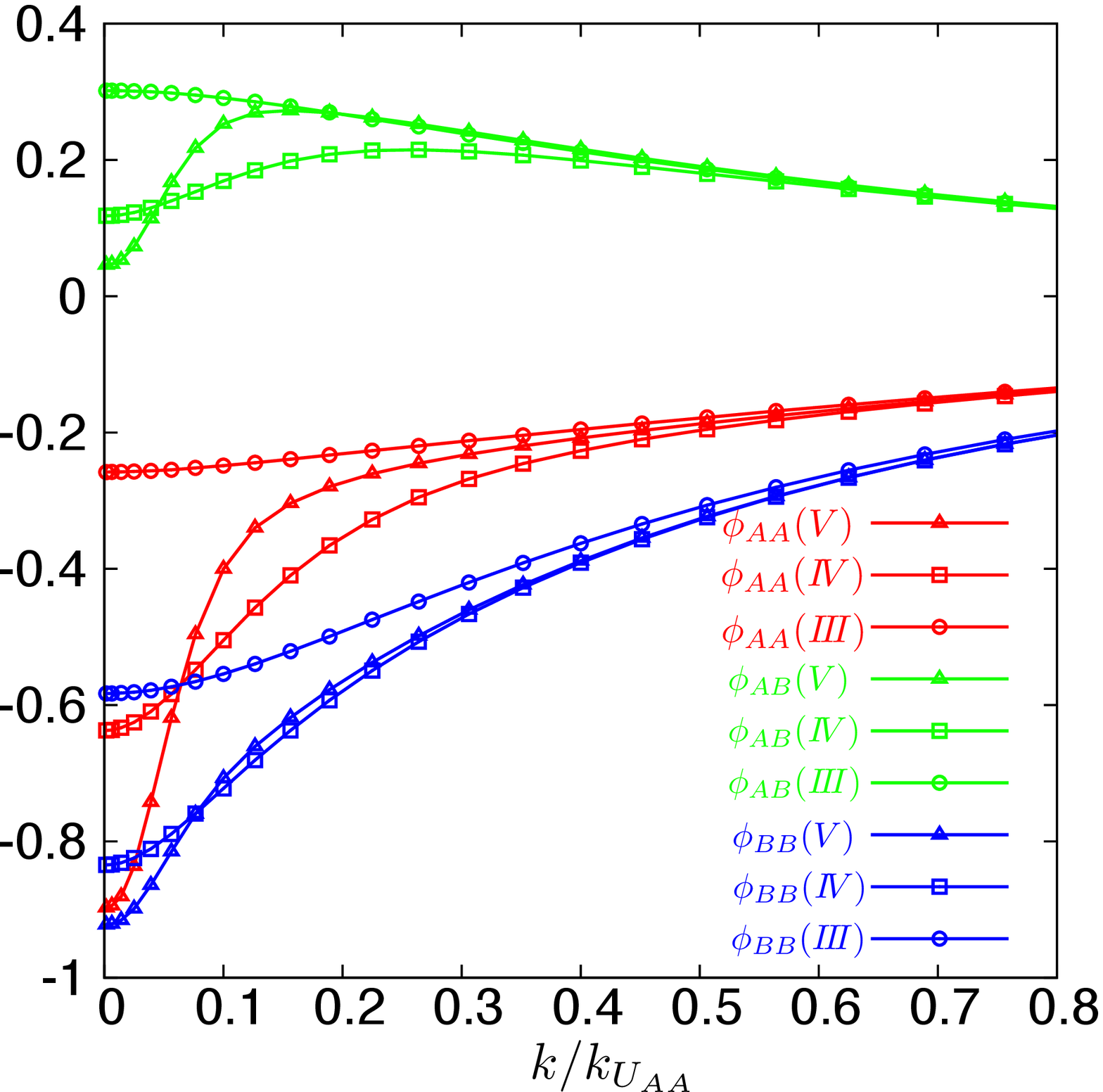}
                \end{center}
\caption{
(Color online)  Variational parameters $\phi_{AA}(k)$ {(red lines)}, $\phi_{AB}(k)$ {(green lines)}, and $\phi_{BB}(k)$ {(blue lines)} given by the case $m_{A}:m_{B}=4:1$, $n_{A}:n_{B}=1:1$, $\tilde{k}_{\rm{c}}=10$, and $U_{AB}= -0.95\sqrt{U_{AA}U_{BB}}$ with conditions $(I\hspace{-.1em}I\hspace{-.1em}I)$ (circular-point lines), $(I\hspace{-.1em}V)$ (square-point lines), and $(V)$ (triangular-point lines)}.
\label{fig3}
\end{figure}

\begin{table}[t]
\caption{{{ {$c_{ij}$ and $\alpha$ with various cases.}}}}
\centering
\begin{tabular}{c c c c c c c}
\hline\hline\\[-2.0ex]
{{$\dfrac{|U_{AB}|}{\sqrt{U_{AA}U_{BB}}}$}}&{{Cut off }}&{{Case  }}&{{ $\dfrac{{c_{AA}}}{\sqrt{\delta_{A}}}$  }}&{{ $\dfrac{c_{AB}}{\sqrt{\delta_{A}}}$}}&{{ $\dfrac{{c_{BB}}}{\sqrt{\delta_{A}}}$ }}&{{ $\dfrac{\alpha}{\sqrt{\delta_{A}}}$ }}\\  [2.0ex] 
\hline
\hline
{{0.95}}&{{$\tilde{k}_{\rm{c}}=5$}}&{{ ($A_{+}$)}}&{{$ 4.88 $}}&{{$ -5.09 $}}&{{$ 4.88 $}}&{{$ 20.2 $}}\\[0.2ex] 
&{{}}&{{($B_{+}$)}}&{{$ 19.5  $}}&{{$ -5.10 $}}&{{$1.24 $}}&{{$ 31.3 $}}\\[0.2ex]  
&{{}}&{{($C_{+}$)}}&{{$ 7.79 $}}&{{$-4.08 $}}&{{$ 1.97 $}}&{{$18.1 $ }}\\[0.2ex] 
&{{}}&{{($A_{-}$)}}&{{$ 4.97 $}}&{{$ 5.21 $}}&{{$ 4.97 $}}&{{$ -0.486 $ }}\\[0.2ex] 
&{{}}&{{($B_{-}$) }}&{{$ 19.9 $}}&{{$ 5.22 $}}&{{$1.26  $}}&{{$ 10.6 $}}\\[0.2ex]  
&{{}}&{{($C_{-}$)}}&{{$ 7.90 $}}&{{$ 4.15 $}}&{{$ 1.99 $}}&{{$1.58 $}}\\[0.2ex]  
\hline
\hline
{{0.95}}&{{$\tilde{k}_{\rm{c}}=10$}}&{{ ($A_{+}$)}}&{{$ 11.7 $ }}&{{$ -12.2 $}}&{{$ 11.7 $}}&{{$ 49.1 $}}\\[0.2ex] 
&{{}}&{{ ($B_{+}$)}}&{{$ 46.7 $}}&{{$ -12.2 $}}&{{$ 2.98 $}}&{{$ 76.0 $ }}\\[0.2ex] 
&{{}}&{{ ($C_{+}$)}}&{{$ 18.8 $}}&{{$ -9.82 $}}&{{$ 4.74 $}}&{{$ 44.0 $ }}\\[0.2ex]  
&{{}}&{{ ($A_{-}$)}}&{{$ 12.3 $}}&{{$ 12.9 $}}&{{$ 12.3 $}}&{{$ -1.18 $  }}\\[0.2ex] 
&{{}}&{{($B_{-}$)}}&{{$ 49.0 $}}&{{$ 12.9 $}}&{{$ 3.11 $}}&{{$ 25.7 $ }}\\[0.2ex]  
&{{}}&{{ ($C_{-}$)}}&{{$ 19.5 $}}&{{$ 10.2 $}}&{{$ 4.91 $}}&{{$ 3.84 $ }}\\[0.2ex]  
\hline
\hline
{{0.98}}&{{$\tilde{k}_{\rm{c}}=5$}}&{{ ($A_{+}$)}}&{{$ 5.21 $}}&{{$ -5.27 $ }}&{{$ 5.21 $}}&{{$ 21.2 $}}\\[0.2ex] 
&{{}}&{{ ($B_{+}$)}}&{{$ 20.8 $}}&{{$ -5.27 $}}&{{$ 1.32 $}}&{{$ 33.0 $ }}\\[0.2ex] 
&{{}}&{{ ($C_{+}$)}}&{{$ 8.31 $}}&{{$ -4.21 $}}&{{$ 2.10 $}}&{{$ 19.0 $ }}\\[0.2ex]  
&{{}}&{{ ($A_{-}$)}}&{{$ 5.30 $}}&{{$ 5.40 $}}&{{$ 5.30 $}}&{{$ -0.183 $  }}\\[0.2ex] 
&{{}}&{{($B_{-}$)}}&{{$ 21.2 $}}&{{$ 5.40 $}}&{{$ 1.34 $}}&{{$ 11.6 $ }}\\[0.2ex]  
&{{}}&{{ ($C_{-}$)}}&{{$ 8.43 $}}&{{$ 4.30 $}}&{{$ 2.12 $}}&{{$ 1.95 $ }}\\[0.2ex]  
\hline
\hline
\end{tabular}
\label{cij}
\end{table}

Figures \ref{fig2} and \ref{fig3} show $\phi_{AA}(k)$ (red lines), $\phi_{BB}(k)$ (blue lines) and $\phi_{AB}(k)$ (green lines) for cases ($C_{\pm}$) with conditions $(I\hspace{-.1em}I\hspace{-.1em}I)$ (circular-point lines), $(I\hspace{-.1em}V)$ (square-point lines), and $(V)$ (triangular-point lines) for the positive and negative signs of $U_{AB}$, respectively.
We see from these figures that the sign of $\phi_{AB}$ directly corresponds to the sign of $U_{AB}$.
Indeed, we have numerically confirmed that $\phi_{AA}({{U_{AB}}})=\phi_{AA}({{-U_{AB}}})$, $\phi_{BB}({{U_{AB}}})=\phi_{BB}({{-U_{AB}}})$, and $\phi_{AB}({{U_{AB}}})=-\phi_{AB}({{-U_{AB}}})$ for condition $(I\hspace{-.1em}I)$.
On the other hand, the amplitudes of the variational parameters in Fig.\ \ref{fig2} are slightly different from the ones in Fig.\ \ref{fig3} due to the presence of the 2- and 3/2-body processes.

We notice that $\phi_{AB}(k\to0)$ for $(I\hspace{-.1em}V)$ and $(V)$ is suppressed compared with that for $(I\hspace{-.1em}I\hspace{-.1em}I)$.
On the other hand, $\phi_{AA}(k\to0)$ and $\phi_{BB}(k\to0)$ for $(V)$ approach $-1$ and are enhanced compared with those for $(I\hspace{-.1em}V)$.
Therefore, (i) pair excitations of different low-lying particles are suppressed by incorporating $3/2$-body processes and (ii) pair excitations of the same low-lying particles are enhanced by the new variational parameters $w_{ABA}$ and $w_{BAB}$.
In this sense, the behaviors of $\phi_{AA}(k)$, $\phi_{AB}(k)$ and $\phi_{BB}(k)$ with condition $(V)$ seem to approach those of the variational parameters with condition $(I\hspace{-.1em}I)$.

Finally, we discuss the correction to the stability condition given by the inequality in Eq.\ (\ref{stab}).
Table \ref{cij} shows $c_{AA}$, $c_{AB}$, $c_{BB}$, and $\alpha$ obtained by numerical calculation based on the ground state with condition $(V)$.
As shown in the table, all the corrections are of order $\sqrt{\delta_{A}}$, which mainly originate from the terms related to $\rho_{AB}(k)$ and $F_{AB}(k)$ in Eqs.\ (\ref{coeff1}) and (\ref{coeff2}). The $3/2$-body processes also give $O(\delta_{A})$ contributions.
In the case of $U_{AB}>0$, we see from the table that $\alpha$ is always positive; thus, the stable regions of the miscible states seem to be extended.
Indeed, we checked that our numerical calculation converged even when $U_{AB}=\sqrt{U_{AA}U_{BB}}$.
On the other hand, in the case of $U_{AB}<0$, $\alpha$ may become negative, especially when $\bar{n}_{A}:\bar{n}_{B}=m_{A}:m_{B}=1:1$.
Under such a condition, we have numerically found that the ground-state energies increased and becomes divergent as the iterative calculations proceeded and the self-consistent calculations became unstable in the range of $U_{AB}\lesssim -0.985\sqrt{U_{AA}U_{BB}}$.
It is difficult to determine the critical point in detail numerically because it is not until we succeed in the self-consistent calculation that we can calculate $\alpha$.
However, these results indicate that many-body effects may change the stable regions of miscible states.

\section{Summary}
We have constructed the ground state for an $M$-component BEC on the basis of self-consistent variational parameters by incorporating the mean-field 2-body processes and dynamical 3/2-body processes between different particles. 
We have numerically shown that $3/2$-body processes lower the ground-state energies, and their contributions are comparable to those of $2$-body processes for $2$-component systems with various masses and particle numbers.
From these results, we suggest that the dynamical interaction processes between condensates and non-condensates such as $NC_{i}+NC_{j}\leftrightarrow C_{i}+NC_{j}$ exist and may yield a comparable contribution to the interaction processes to that of $NC_{i}+NC_{j}\leftrightarrow NC_{i}+NC_{j}$ in mixed BECs. 
We have also reconsidered the stability condition for $2$-component miscible states on the basis of a new ground-state wave function and obtained the new inequality $U^{2}_{AB}/U_{AA}U_{BB}<1+\alpha$, where $\alpha$ originates from $2$-body and $3/2$-body processes.
Since $\alpha$ is on the order of the square root of the coupling constant, many-body effects cannot be neglected, especially in the case that critical points are investigated experimentally in systems with strong coupling.

According to the study of single-component systems\cite{kita3/2}, 3/2-body processes (i) characterize the qualitative difference between one-particle excitations and collective modes of BECs by giving rise to the finite widths of single-particle spectra even for $k\to 0$ and (ii) play a role in maintaining the macroscopic coherence of BECs in equilibrium. 
In this sense, to reveal the nature of microscopic 3/2-body processes in mixed BECs, microscopic physical quantities such as single-particle spectra should be studied on the basis of $\ket{\Phi}$ obtained by this study.

From the viewpoint of constructing the ground state of BECs, the variational method in the present study incorporating the many-body effect is expected to be  directly and self-consistently applied to other various systems, such as spinor BEC \cite{Spinor} and boson-fermion mixtures \cite{BoseFermi} by considering the contributions of spin-flip processes and boson-fermion processes, respectively.
In addition, the ground state of a BEC trapped by a potential and the density matrix for finite-temperature systems have not been constructed.
In this context, this self-consistent variational method with 2- and 3/2-body processes will give new prospects for studying BEC.

\end{document}